\documentclass[usegraphicx,usenatbib]{mn2e}

\def\simlt{\lower.5ex\hbox{$\; \buildrel < \over \sim \;$}}
\def\simgt{\lower.5ex\hbox{$\; \buildrel > \over \sim \;$}}
\def\simpropto{\lower.2ex\hbox{$\; \buildrel \propto \over \sim \;$}}
\begin{document}

\title[CMB Anomalies viewed via Gumbel Statistics]{Cosmic microwave background anomalies viewed via Gumbel Statistics}

\author[Mikelsons, Silk \& Zuntz]{Gatis Mikelsons, Joseph Silk \& Joe Zuntz}



\def\ellmax{$\ell_{\textrm{max}}$}
\def\fnl  { \ensuremath{ f_{\mathrm{NL}} } }
\def\ellmax  { \ensuremath{\ell_{\mathrm{max}} } }

\maketitle
\begin{abstract}
We describe and discuss the application of Gumbel statistics, which model extreme events, to WMAP 5-year measurements of the 
cosmic microwave background.  We find that temperature extrema of the CMB are well modelled by the Gumbel formalism and describe 
tests for Gaussianity that the approach can provide.  Comparison to simulations reveals Gumbel statistics to have only weak 
discriminatory power for the conventional statistic: $\fnl<1000$, though it may probe other regimes of non-Gaussianity. Tests 
based on hemispheric cuts reveal interesting alignment with other reported CMB anomalies. The approach has the advantage of model 
independence and may find further utility with smaller scale data.
\end{abstract}


\section{Introduction}

Although the primary source of information in the cosmic microwave background (CMB) is the angular power spectrum, tests of 
other statistics provide an important consistency check for cosmological models.  The key assumptions of Gaussianity and isotropy 
of the CMB, particularly of the WMAP data \citep{CMBsource}, have both been challenged by more complex tests. The isotropy of the 
CMB has been strongly challenged with respect to power distribution \citep{hansen} and multipole alignment \citep{land, copi}, and 
some evidence for non-Gaussianty has been noted among many statistical tests \citep{vielva, eriksen, mukherjee, mcewen}.  
Many such tests use higher order statistics, moving beyond the correlation of pairs of sky pixels and using more complex 
comparisons. The statistics of CMB extrema have previously been studied by looking at the statistics of hot and cold spots 
\citep{larson,hou}; here we address the inverse question: given patches of fixed area, what are the statistics of the extremes within 
those patches? We develop and demonstrate the use of a new statistical approach, based on the statistics of extreme events. 
We find that the test has weak discriminatory power but shows some suggestion of non-Gaussianity at low $\ell$.

Gumbel statistics describe the behaviour of sample extrema in the same way that Gaussian statistics are used to describe the 
behaviour of sample means. For example, if we draw $n$ independent values from a Gaussian distribution $N(\mu,\sigma)$, the average of those 
values will also be Gaussian-distributed with mean $\mu$ and standard deviation $\sigma/\sqrt{n}$. Similarly, the 
maximum and minimum of that sample will, in the limit, follow a Gumbel distribution \citep{gmb1}.

Just as the central limit theorem implies that sample means from any distribution will, in the limit, tend to a Gaussian profile, 
all sample extrema will tend to a Gumbel distribution for sufficiently large sample size. The distribution is also known in 
literature as the von Mises family of distributions \citep{gmb1}. 

Gumbel statistics have found application in a number of fields, primarily financial and meteorological. In this paper we apply 
Gumbel statistics to CMB-type maps and hence discuss cosmological applications of Gumbel distributions. Section 
\ref{sec:distribution} presents the mathematical basis of Gumbel statistics. In Section \ref{sec:cmb} we apply the formalism to 
the CMB and simulated maps to demonstrate our method. Section \ref{sec:cmb_ng} compares the findings for the CMB to those for 
simulated Gaussian maps. Sections \ref{sec:ng} applies Gumbel statistics to searching for potential non-Gaussianities in simulated 
maps.

\section{Gumbel Distributions}\label{sec:distribution}

\begin{figure}
\includegraphics[width=\columnwidth]{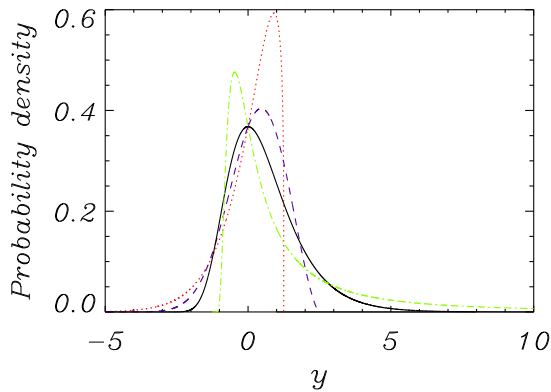}
\caption{Selection of Gumbel probability density functions for sample maxima. Plotted values of $\gamma$ are: $-0.8$ (dot), $-0.4$ 
(dash), $0$ (solid), $0.8$ (dash-dot).\label{gpdf}}
\end{figure}

The most general Gumbel profile for sample maxima is given by the cumulative distribution function of the form \citep{gmb1, gmb2}:
\begin{equation}
C_{\gamma}(y)=exp(-(1+{\gamma}y)^{-1/\gamma})
\label{defcdf} 
\end{equation} 

where $y$ is related to the variable value $x$ via scale and intercept parameters: $y=(x-a)/b$. In the context of the CMB $x$ is the maximum temperature of a pixel in 
a patch.  The quantity $\gamma$ is a shape 
parameter that can take any real value, depending on the underlying distribution. Figure \ref{gpdf} plots the probability density 
functions, corresponding to Equation \ref{defcdf}, using a selection of $\gamma$ values. A similar family of distributions exists 
for the sample minimum, and we shall refer to both collectively as `Gumbel statistics'.  

The value of $\gamma$ for many analytic underlying distributions can be calculated rigorously \citep{gmb1, gmb2}. For 
example, a sample of identically distributed Gaussian variables can be shown to yield $\gamma$ equal to zero for both maxima and 
minima. In general, though, the $\gamma$ one calculates will be different for the maxima and minima limit distributions.  The distribution of the CMB, on the other hand,
is a more complicated multi-dimensional Gaussian, and there is no analytic prediction for its Gumbel statistics.

We can, though, estimate the $\gamma$ parameter by repeatedly taking a ``sample of samples'' from a distribution - taking $n$ sets of 
values from the distribution, each containing $m$ measurements, and recording the maximum in each set.  When the number $n$ of 
such maxima is large enough, the corresponding density function can be fitted to a Gumbel form.  The $\gamma$ for the underlying 
distribution will be reflected accurately in the data, provided that $m$ is large enough for an analogy of the central limit 
theorem to hold, and we will be able to estimate its value accurately if $n$ is large enough.

\section{Gumbel statistics of CMB maps}\label{sec:cmb}

To study the extreme statistics of CMB-type maps, we make use of randomly placed circular regions (`patches'). We position a 
number of such patches on the CMB sky and record the maximum and minimum temperatures within them (Figure \ref{600_4deg_proc}). For 
Gumbel statistics to be applicable to this set of extremes, the patches must be large enough to contain sufficient independent 
temperatures for the analogy to the central limit theorem to hold. This is a function of the pixel and patch size, as well as the 
map resolution. 
We take as our CMB data the five-year Internal Linear Combination (ILC) map \citep{CMBsource}. An upper limit to the Fourier resolution (maximum $\ell$ value) credibly usable in the ILC map is set by foreground emission, 
so that at $\ellmax>25$ the true CMB signal becomes progressively contaminated \citep{CMBsource}. Another constraint is having only 
a finite CMB sky to sample, as a result of which our sets of samples will tend to overlap if patch sizes are too large. The ideal 
patch size would be large enough to allow sufficient variability, yet small enough to prevent overlapping. In practice, patches of 
angular radius between $2^\circ$ and $8^\circ$ fulfill both these conditions.  

The ILC map is certainly not ideal: we are forced to use patches on scales small enough that foregrounds can contaminate them.  We avoid this by removing high spatial frequency modes from the map, but this will reduce the detectability of any real signals that are present on smaller and intermediate scales.  A full analysis could employ data from the high-signal frequency bands and a much more conservative mask, but the ILC map serves to illustrate the process and perform basic tests.

\subsection{Methodology}

We test whether Gumbel statistics are applicable to CMB-type data by computing $\gamma$ for the CMB and simulated Gaussian maps. 
Since we will later examine the use of Gumbel statistics in detecting non-Gaussianity (Section \ref{sec:ng}), we also test 
simulated maps with a varied non-Gaussian signal generated using the method of Liguori et. al. \citep{liguori}.

\begin{figure}
\includegraphics[angle=90,width=\columnwidth]{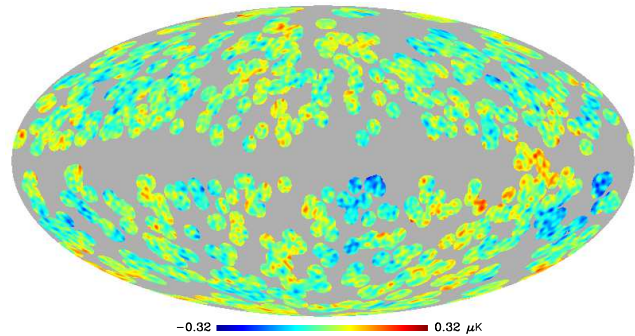}
\caption{Random placement of 600 patches of $4^\circ$ radius over a CMB map with the Processing mask 
applied.\label{600_4deg_proc}}
\end{figure}

Our algorithm is as follows. To prevent the worst contamination from galactic foreground emission, we first mask the data using the 
processing mask supplied by the WMAP team \citep{CMBsource}, which covers 6\% of the sky. We then randomly place patches of fixed angular radius 
over the remaining map (Figure \ref{600_4deg_proc}) and record the maximum/minimum temperature values in each region enclosed. The 
sets of maxima and minima are then converted into CDF profiles to which a Gumbel distribution (e.g. Equation (\ref{defcdf}) for 
the maxima) can be fitted. This way, one obtains the Gumbel parameters $a$, $b$ and $\gamma$ for the map in question. Parameter 
errors are obtained, using a simple Monte Carlo process over the whole algorithm. Error bars drawn in subsequent plots will 
correspond to $1\sigma$ in total width.

Our simulated maps are realisations of a concordance $\Lambda CDM$ 
cosmology with WMAP standard cosmological parameters \citep{CMBparam}. Our non-Gaussian maps are generated using the same power 
spectrum but have a non-Gaussian component with a tunable value of\fnl.

Analysis on simulated maps is done in the same manner as for the ILC, including the masking of the sky. The only difference is 
introduced in the Monte Carlo algorithm, where we generate a new simulated map each time the random placement of patches is rerun. 
By doing so we obtain estimates of Gumbel parameters for all maps with a particular power spectrum.

\subsection{Results}

\subsubsection{Fit Quality}

\begin{figure}
\includegraphics[width=\columnwidth]{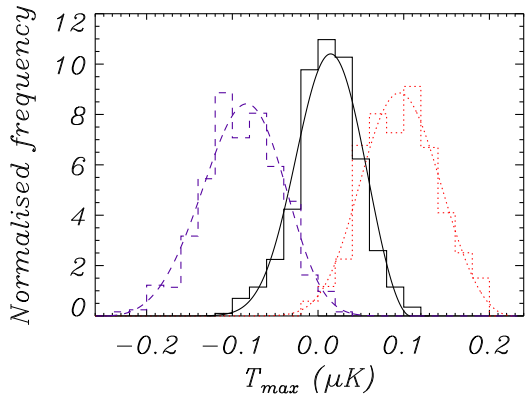}
\caption{Sample Gumbel fits to the CMB data. Fitted Gumbel PDF's are shown superimposed on the temperature histograms, normalised 
to unity. The data correspond to: maxima fit using 600 $4^\circ$ patches at $\ellmax=96$ (dot), maxima fit using 1000 $2^\circ$ 
patches at $\ellmax=16$ (solid), minima fit using 600 $4^\circ$ patches at $\ellmax=96$ (dash).\label{threehist}}
\end{figure}

For both real and simulated data, the collected sets of maxima and minima are very well fit by the Gumbel form with parameters $a,b,\gamma$.  Figure 
\ref{threehist} shows three sample fits with the ILC data for both maxima and minima. Evidently, in all cases the fits are close 
to ideal.  This was also the case for simulated maps, both Gaussian and non-Gaussian (examined up to $\fnl=4000$).  The Gumbel 
$\gamma$ parameter seems therefore to be a well-defined statistic characterizing a CMB-type map at a given resolution.

\subsubsection{Stability of the Gumbel $\gamma$ parameter}

We next investigate the variation in $\gamma$ with sample size. We expect $\gamma$ to be independent of the sample (patch) size, 
once a critical value is reached, since one would then keep approaching the limiting distribution with increasing accuracy. 

\begin{figure}
\includegraphics[width=\columnwidth]{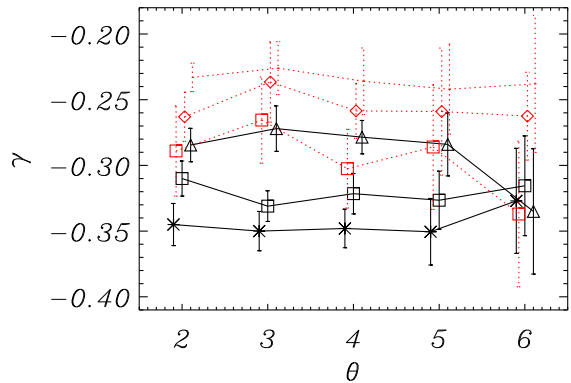}
\caption{Gumbel $\gamma$ versus patch radius plotted for the maxima of the CMB map (solid) and simulated Gaussian maps (dot). The 
various values of \ellmax correspond to: $\ellmax=16$ (star), $\ellmax=32$ (square), $\ellmax=96$ (triangle), $\ellmax=128$ 
(diamond), $\ellmax=500$ (none)\label{thwoscatters}}
\end{figure}

We test the stability of $\gamma$ by repeating the fitting procedure for varying patch size, using in each case the maximum number 
of patches without oversampling. Results are shown in figure \ref{thwoscatters} for patches varied between $2^\circ$ to $6^\circ$ in radius, with total number 
of patches ranging between $1000$ and $100$, respectively. Only results for maxima are displayed, since the minima fits lead to similar conclusions. 
Since they are free of foregrounds, we examine the simulated data over a greater range of \ellmax.

As can be seen from Figure \ref{thwoscatters}, the results are consistent with $\gamma$ remaining constant over the patch size for 
both the ILC and simulated Gaussian maps, smoothed to a variety of \ellmax. This result was also seen to hold for simulated non-Gaussian 
maps. We conclude that the Gumbel treatment is valid for a range of simulated maps, with non-Gaussianity spanning between $\fnl=0$ 
and $\fnl=4000$, and for the ILC map.

\subsubsection{Variation in the values of $\gamma$}

Over the range of maps studied, for both maxima and minima, the Gumbel treatment invariably yields negative values of $\gamma$, 
corresponding to the Weibull domain in extreme value theory \citep{gmb2}.  For this domain, the PDFs of temperature maxima are 
limited on the right (Figure \ref{gpdf}), whereas the ones for temperature minima are limited on the left; i.e. in both cases the 
tail of interest has an upper bound.

It is also observed that the $\gamma$ values for maxima tend to be generally lower and vary more with \ellmax than for the minima. 
In the case of the ILC, Gumbel statistics differentiate positive and negative temperature fluctuations within the data by a clear spread (Figure \ref{band}). 

This result is not true for simulated Gaussian maps, where the values for maxima closely mirror those for the minima. This is 
expected, given a map generated to be symmetric about the zero temperature.

\section{Comparison with Gaussian simulations}\label{sec:cmb_ng}

A pertinent question to investigate is whether or not Gumbel statistics of the CMB differ from statistics collected using 
simulated Gaussian maps. Any notable difference would be evidence against the 5-year ILC map being a purely Gaussian data set.

To study possible differences, we collect values of $\gamma$ from the ILC and from simulated Gaussian maps using a fixed patch 
radius ($\theta=4^\circ$) and examine their variation with \ellmax. We present data for four different Fourier resolutions: 
$\ellmax=16, 32, 64$ and $96$. Increasing \ellmax further would imply using the data at much higher accuracy than recommended by 
the WMAP team and is also found to add no significant variation. Using lower \ellmax, on the other hand, would reduce the number 
of independent temperatures inside a patch, invalidating the Gumbel approach.

\begin{figure}
\includegraphics[width=\columnwidth]{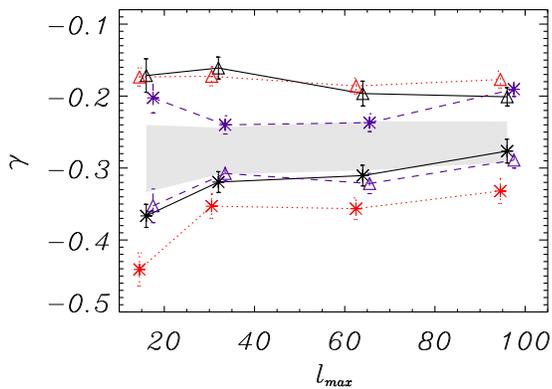}
\caption{Gumbel $\gamma$ versus \ellmax, plotted for the CMB and simulated Gaussian maps. The data correspond to: CMB full sky 
(solid), CMB North hemisphere (dash), CMB South hemisphere (dot). Both maxima fits (star) and minima fits (triangle) are given. 
The grey band corresponds to the full-sky results (maxima, minima) for simulated Gaussian maps, averaged over 100 realisations. 
\label{band}}
\end{figure}

Figure \ref{band} (solid line) shows results for the CMB, which can be compared with the Gaussian data (grey band), collected and 
averaged over 100 simulated maps. The band is then an indication of where data for a single Gaussian realisation would be expected 
to lie. Separate results for the CMB North (dash) and South (dot) hemispheres are also plotted. 

As seen, data for the CMB display slight variation with \ellmax, though not of high statistical significance. More importantly, 
the CMB data show a spread of several $\sigma$ between the maxima and minima which is not present in the simulated Gaussian results.  

\begin{figure}
\includegraphics[width=\columnwidth]{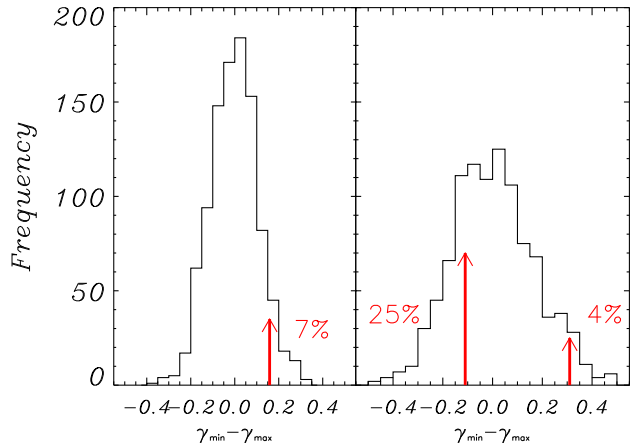}
\caption{Left: histogram showing the spread of $\gamma_{min}-\gamma_{max}$ for 1000 simulated full-sky Gaussian maps, examined at 
$\ellmax=32$. The arrow indicates the CMB value. Right: histogram showing the range of $\gamma_{min}-\gamma_{max}$ for 1000 
simulated Gaussian hemispheres ($\ellmax=32$). The arrows indicate the range of CMB values, as seen in Figure 
\ref{diffmap}.\label{thists}}
\end{figure}

To quantify the likelihood of this separation between maxima and minima, we generate 1000 Gaussian maps and record values of $\gamma_{min}-\gamma_{max}$, collected at 
$\ellmax=32$. The resulting histogram is shown in Figure \ref{thists} (left), with the arrow indicating the corresponding spread in 
the CMB. Evidently, on this criterion the ILC map is still consistent with being a Gaussian set.

On further investigation, it is found that the spread in $\gamma$ within the CMB varies considerably with direction. Rerunning the 
same analysis with the ILC map split into North and South hemispheres we find that values of $\gamma$ change their magnitude by 
several $\sigma$, and the spread between maxima and minima is actually reversed (Figure \ref{band}, dot, dash). The result for the 
whole sky (solid) masks considerable anisotropy within the data set. Variations of this kind are 
not found in Gaussian maps, when averaged over, though individual realisations with similar North-South asymmetry were 
encountered.

\begin{figure}
\includegraphics[width=\columnwidth]{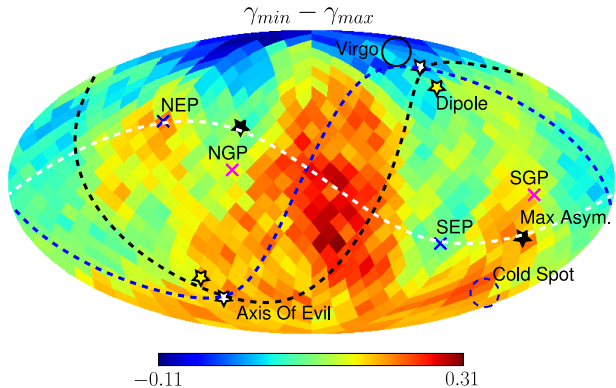}
\caption{A map recording $\gamma_{min}-\gamma_{max}$ for the CMB ($\ellmax=32$), as measured using hemispheres of different 
angular orientation. The pixel position corresponds to the orientation of the rotated North axis. Angular positions of the 
Ecliptic and Galactic poles are shown superimposed (NEP, SEP, NGP, SGP) as is the location of the Virgo cluster and the
orientation of the CMB dipole \citep{CMBparam}. Three known CMB anomalies are also indicated: the cut of maximum
hemispheric power asymmetry \citep{eriksen}, the Axis of Evil \citep{land_early}, and the Cold Spot \citep{vielva}
}

\label{diffmap}
\end{figure}

To survey this anisotropy within the CMB, we rotate the North-South axis to varying orientations and repeat the analysis for the corresponding `North' hemisphere only. Results are shown in Figure \ref{diffmap}. Here, the spread 
in $\gamma$ for each rotated hemisphere ($\gamma_{min}-\gamma_{max}$ measured at $\ell=32$) is shown as the value for the pixel, 
located along the direction of the rotated `North' axis. 

The value of $\gamma_{min}-\gamma_{max}$ in a hemisphere is thus seen to vary between $-0.11$ and $0.31$, depending on 
orientation. To compare with Gaussian maps, we record the same statistic for 1000 simulated hemispheres, as shown in Figure 
\ref{thists} (right). We thus find that the variation in $\gamma_{min}-\gamma_{max}$ within the CMB is of magnitude comparable to 
simulated Gaussian maps, though the strongest signal would only be reproduced in $<10\%$ of the simulated maps.

Figure \ref{diffmap} also plots well-known CMB anomalies that have previously been reported \citep{land_early, eriksen, vielva}. 
The alignment which maximizes the Gumbel discriminant is seen to be, like the Axis of Evil \citep{land_early}, close to the equator of another previously reported anomaly - the 
anisotropic division of power between the ecliptic hemispheres \citep{eriksen}.

In summary, having examined the Gumbel statistics of the CMB, we can report considerable variety in the values of $\gamma$ within 
the data set. The data are consistent with Gaussianity, though the observed signal is only reproduced in $<10\%$ of the simulated 
Gaussian maps.

\section{Non-Gaussianity limits}\label{sec:ng}

Having established the validity of the Gumbel approach as applied to the CMB and simulated data (Section \ref{sec:cmb}), we probe 
further its usefulness in detecting non-Gaussian CMB signals, generated by quadratic perturbations in the primordial potential 
field,\fnl. For a review of previous work, see \citet{yadav, vielva, mukherjee}.

\begin{figure}
\includegraphics[width=\columnwidth]{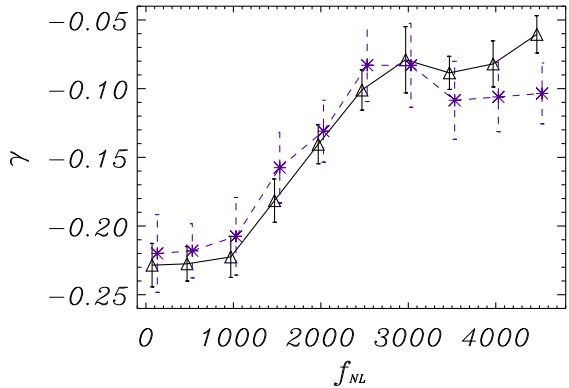}
\caption{Gumbel $\gamma$ as obtained for maxima of simulated non-Gaussian maps ($\ellmax=500$) of varying\fnl. Results are 
plotted for patch sizes $2^\circ$ (star) and $4^\circ$ (diamond).\label{gfnl}}
\end{figure}

To begin, we examine the variation of $\gamma$ with\fnl using high-resolution simulated maps with $\ellmax=500$. We vary\fnl 
over the range of $0$ to $4500$, a region over which $\gamma$ has been tested to be a reliable parameter. Figure \ref{gfnl} plots 
the results.

We observe a monotonic increase in $\gamma$ as \fnl is varied from 1000 to 3000 after which point the value seems to stabilise. 
Similar though less steep monotonic increase is found when plotting values of $\gamma$ for Gumbel minima.

The size of error bars on the $\gamma$ parameter in Figure \ref{gfnl} constrain the lowest signal likely to be detected by this 
method to approximately $\fnl=\sim2000$. The procedure seems to afford no way of obtaining more precise $\gamma$ estimates, which would allow us to detect lower $\fnl$.

The ILC exhibits some divergence from the Gaussian simulations at about the $10\%$ level (see above).  There are three possible 
explanations for this.  One possibility is residual impact from foregrounds, even at these large scales. Secondly, the 
non-Gaussian signature might be of a form not well described by \fnl. The third possibility is a simple statistical fluctuation.

\section{Conclusion}
We have introduced the statistics of sample extremes, Gumbel statistics, to the study of CMB maps. We have shown the statistics to 
provide a good description of the WMAP data set despite the highly correlated fluctuations therein.

We have investigated the use of Gumbel statistics in non-Gaussianity detection, the principal advantage of our method being its 
lack of restrictions as to the underlying statistics of the CMB. The simplest methods, based on Gumbel statistics, are 
unlikely to detect any $\fnl<1000$.  There are weak hints of non-Gaussianity at low $\ell$, as we can only reproduce the 
observed statistics some $10\%$ of the time in Gaussian realisations, but no hard evidence. We also find that there is a preferred hemispheric cut that 
optimises the Gumbel discriminant. We emphasize that\fnl is a limited and non-generic description of non-Gaussianity. Gumbel 
statistics can provide a potential probe of generic non-Gaussianity in CMB data sets.

\section{Acknowledgements}
We are grateful to M. Liguori and collaborators for providing us with non-Gaussian CMB simulations generated using their code and 
algorithm, and to Joanna Dunkley for useful discussion. Our calculations made use of the HEALPIX package \citep{healpix}. JZ is funded by an STFC rolling grant.

\end{document}